# Lifetime of the singly charged $^{229}$Th nuclear isomer

Y. Shigekawa[1,2,*], A. Yamaguchi[3,4], K. Tokoi[5], N. Sato[1], H. Kikunaga[6], K. Shirasaki[7,8], Y. Kasamatsu[5], M. Wada[9,10], and H. Haba[1]

[1]*Nishina Center for Accelerator-Based Science, RIKEN, Wako, Saitama 351-0198, Japan*

[2]*Institute of Pure and Applied Sciences, University of Tsukuba, Tsukuba, Ibaraki 305-8577, Japan*

[3]*Space-Time Engineering Research Team, RIKEN Center for Advanced Photonics, RIKEN, Wako, Saitama 351-0198, Japan*

[4]*PRESTO, Japan Science and Technology Agency, Kawaguchi-shi, Saitama 332-0012, Japan*

[5]*Graduate School of Science, The University of Osaka, Toyonaka, Osaka 560-0043, Japan*

[6]*Research Center for Accelerator and Radioisotope Science, Tohoku University, Sendai, Miyagi 982-0826, Japan*

[7]*Institute for Materials Research, Tohoku University, Sendai, Miyagi 980-8577, Japan*

[8]*Institute for Radiation Science, The University of Osaka, Toyonaka, Osaka 560-0043, Japan*

[9]*Wako Nuclear Science Center, Institute of Particle and Nuclear Studies, High Energy Accelerator Research Organization, Wako, Saitama 351-0198, Japan*

[10]*Advanced Energy Science and Technology GuangDong Laboratory, Huizhou 516007, China*

**Abstract**

The nucleus of thorium-229 ($^{229}$Th) has an exceptionally low-energy isomeric state ($^{229m}$Th). Because its excitation energy is close to the energy levels of valence electrons, the lifetime of $^{229m}$Th varies substantially depending on its electronic state. Although internal conversion and radiative decay were recently observed, the electronic bridge decay of $^{229m}$Th, a higher-order decay process through an electronic transition, has not yet been confirmed. A promising candidate to search for this decay channel is singly charged $^{229m}Th^{+}$ in the electronic ground state. Here we produce $^{229m}Th^{+}$ by a charge exchange reaction in an ion trap and detect the isomers through measuring the electrons emitted from internal conversion processes. We determined the half-life of $^{229m}Th^{+}$ to be 0.46(8) s. Our result differs by several orders of magnitude from the half-lives of internal conversion and radiative decay, indirectly suggesting the existence of the electronic bridge decay of $^{229m}Th$. This will enable the direct observation and further investigation of the electronic bridge process, which will contribute to understanding nuclear–electron interactions and accelerating nuclear deexcitation in the operation of a thorium-based nuclear clock.

*yshigekawa@ied.tsukuba.ac.jp

**Main text**

The nuclear decay half-life is normally constant regardless of the external environment because the contribution of valence electrons to nuclear decay is negligibly small; however, the half-lives of some nuclei such as $^{99m}Tc$ and $^{235m}U$, which undergo low-energy nuclear transitions in their deexcitation processes, are known to change slightly depending on the electronic environment[1]. The most extreme case is the first excited state of the $^{229}Th$ nucleus ($^{229m}Th$), which has the lowest excitation energy of all known nuclei. The excitation energy of $^{229m}Th$ above the ground state ($^{229g}Th$) is 8.36 eV[2–4], which allows its decay mode to vary depending on the electronic states of valence electrons and leads to significant half-life variations[5]. Previous studies have observed the internal conversion (IC) decay of $^{229m}Th$ with a half-life of 7(1) μs for neutral $^{229}Th$ atoms on a microchannel plate (MCP) detector[6,7]. This short-lived IC decay was suppressed for $^{229m}Th$-doped fluoride crystals[2–4,8–10] and $^{229m}ThF_4$[11], allowing the observation of γ-ray emission (radiative decay). The half-lives were measured to be 100–670 s, which were more than seven orders of magnitude longer than the IC decay. The half-lives for triply and doubly charged $^{229m}Th$ ions were also found to be long ($1400^{+600}_{-300}$ s[12] and >60 s[6,13], respectively). The narrow natural linewidth of the nuclear transition, resulting from the long isomer lifetime where the IC decay is suppressed, is advantageous toward a nuclear clock with unprecedented accuracy[14–16].

In addition to IC and radiative decay, $^{229m}$Th is predicted to decay by the electronic bridge (EB)[17,18], in which nuclear excitation energy is transferred to the transition of an atomic electron, accompanied by the emission of a photon. The EB decay rate of $^{229m}$Th may vary significantly depending on its electron configuration[19–22] and chemical environment. This presents an opportunity to investigate this higher-order nuclear transition process, where nuclear-electron interactions play an important role. However, the EB decay of $^{229m}$Th has yet to be observed.

A candidate for the electronic states where $^{229m}$Th predominantly decays by the EB process is the singly charged $^{229m}$Th ($^{229m}Th^{+}$)[23–26] in the electronic ground state because many unoccupied electronic states exist around 8 eV above the electronic ground state[27,28]. In the previous experiments[6,7], the low-energy ion beams of $^{229m,g}Th^{3+}$ and $^{229m,g}Th^{2+}$ ($^{229m}$Th:$^{229g}$Th$\approx$2%:98%[13,29]) extracted from a $^{233}$U source were transported onto an MCP detector and the IC electrons emitted from $^{229m}$Th neutralized on the MCP surface were detected. However, when $^{229m,g}Th^{+}$ ions were transported, no IC electrons were observed, suggesting that the isomer rapidly decayed during the ion extraction time (~10 ms) and the lifetime of $^{229m}Th^{+}$ was less than 10 ms, which was shorter than theoretically calculated values[23–26]. In these experiments, $^{229m}Th^{+}$ ions were produced while multiply-charged high-energy (84 keV) $^{229m}$Th ions emitted from the $^{233}$U source were decelerated and transported in helium (He) buffer gas (4 kPa). During this process, electron excitations in $^{229m}Th^{+}$ may have occurred and opened the short-lived IC decay process[30], or the collisional broadening of electronic states may have reduced the EB decay lifetime[7,25]. To further understand the decay process of $^{229m}Th^{+}$, it is desirable to produce $^{229m}Th^{+}$ through low-energy collisions under low gas pressure and measure its half-life.

In this study, we produced $^{229m}Th^{+}$ by a charge-exchange reaction between $^{229m}Th^{2+}$ and nitric oxide (NO) gas[31] inside an ion trap. We extracted $^{229m}Th^{+}$ onto an MCP detector and observed the IC electrons emitted from $^{229m}$Th neutralized on the detector. This IC-electron observation enabled us to count the number of $^{229m}Th^{+}$ ions in the ion trap over time and to determine the half-life of $^{229m}Th^{+}$, which indirectly suggested the existence of the EB decay.

## Extraction of $^{229m,g}Th^{2+}$ and production of $^{229m,g}Th^{+}$

Figure 1a shows the setup for extracting $^{229m,g}Th^{2+}$ and producing $^{229m,g}Th^{+}$. First, multiply charged $^{229m,g}Th$ ions recoiling from a $^{233}U$ source (600 kBq; diameter 90 mm; 0.226(2) ppm of $^{232}U$) were thermalized in He buffer gas (1 kPa). Here, $^{229m,g}Th^{3+}$, $^{229m,g}Th^{2+}$, and $^{229m,g}ThO^{2+}$ were mainly produced by charge-exchange and chemical reactions with He and gaseous impurities. $^{233}U$ ions were also produced by sputtering due to recoil ions and α particles. The thermalized ions were pushed by a direct current (DC) gradient (7.6 $Vcm^{-1}$) toward a radio-frequency (RF) carpet[32] and then extracted from the central hole of the RF carpet using the ion-surfing technique[33,34]. The ions were subsequently transported with a quadrupole ion guide into a linear quadrupole ion trap comprising three segments. The ion-trap region was filled with He gas flowing from the RF-carpet gas cell and a mixture of argon (Ar) and NO gases introduced through a variable leak valve (Methods). The Ar gas was used to cool the ions in the ion trap, while the NO gas was used to produce $^{229m,g}Th^{+}$ by the charge-exchange reaction. NO gas was selected because its ionization energy (9.26 eV) is lower than the second ionization energy of Th (12.1(2) eV[27]) and a previous study[31] showed that $Th^{+}$ ions were predominantly produced by the reaction between $Th^{2+}$ and NO gas. The total pressure ($p_{total}$) in the ion-trap region was maintained at 0.40 Pa, and the partial pressures of the He, Ar, and NO gases were $p_{He} \approx 0.16$ Pa, $p_{Ar} \approx 0.24$ Pa, and $p_{NO} < 1\times10^{-5}$ Pa, respectively. In the ion-trap experiments, ions from the RF-carpet gas cell were first loaded into trap segment 1. Ion loading was stopped by switching the DC voltage applied to the $^{233}U$ source from +40 V to −40 V, after which the ions were trapped for a certain period. The ions were then extracted from the trap, sent to a quadrupole mass separator, and transported to the MCP detector.

We first continuously extracted the ions to the MCP detector and obtained the mass-to-charge ratio (*m*/*z*) spectrum shown as the blue curve in Fig. 1b. $^{229m,g}Th^{2+}$ ions were extracted at a rate of ~2.0×10$^4$ ions/s (including ~400 ions/s of $^{229m}Th^{2+}$), calculated using an MCP ion detection efficiency of ~50%. When we loaded ions for 0.5 s and trapped them for 0.1 s at $p_{NO} = 6.1(3)\times10^{-6}$ Pa, we obtained the *m*/*z* spectrum shown as the red curve in Fig. 1b. Here, the number of $^{229m,g}Th^{2+}$ decreased compared to continuous extraction, while that of $^{229m,g}Th^{+}$ increased by a factor of 13. Hence, we were able to produce $^{229m,g}Th^{+}$ by the charge-exchange reaction between $^{229m,g}Th^{2+}$ and NO gas[31] inside the ion trap. We note that the number of $^{229m,g}Th^{+}$ ions directly extracted from the RF carpet gas cell was small (<several ions/s) under the experimental parameters of the main measurements (Method) and almost all of the $^{229m,g}Th^{+}$ ions shown in Fig. 1b were produced in the ion trap.

In this study, we extracted mass-separated $^{229m,g}Th^{2+}$ or $^{229m,g}Th^{+}$ onto the MCP detector, the surface voltage of which was set to −2000 or −35 V. At −2000 V, ion-impact signals could be measured because the ions were accelerated to ~2000 eV, which produced secondary electrons on the MCP detector with high probability. At −35 V, the ions were softly placed on the MCP surface, significantly reducing ion-impact signals and enabling the detection of the IC electrons emitted from $^{229m}Th$[6,7].

**Isomer detection by IC electron measurement**

Previous experiments[7] established the method for measuring the IC electrons of $^{229m}Th$ using ion bunches of $^{229m,g}Th^{2+}$ and $^{229m,g}Th^{3+}$ with bunch widths of ~10 μs, which were short enough to discriminate between IC-electron and ion-impact signals. We performed similar measurements for $^{229m}Th^{2+}$ and $^{229m}Th^{+}$.

We prepared short ion bunches of $^{229m,g}Th^{2+}$ (~320 ions/bunch) and $^{229m,g}Th^{+}$ (~200 ions/bunch) by loading and trapping the ions downstream of trap segment 1 (Methods). First, we measured the ion-impact signals at an MCP surface voltage of −2000 V (blue triangles in Figs. 2a and 2b). Here, the widths of the ion bunches of $^{229m,g}Th^{2+}$ and $^{229m,g}Th^{+}$ are 2 and 3 μs in the full width at half maximum, respectively. Next, we set the MCP surface voltage to −35 V to measure the IC electrons of $^{229m}Th$, where the ion-impact signals were three orders of magnitude lower than those at −2000 V. Both the data of $^{229m,g}Th^{2+}$ and $^{229m,g}Th^{+}$ at −35 V (black circles in Figs. 2a and 2b) show decay components with lifetimes clearly longer than those at −2000 V.

We discuss the origins of these long-lived decay components observed at −35 V to show that they correspond to the IC electrons of $^{229m}Th$. As described in Ref. [6], the potential background sources are grouped into four categories: (A) background by the kinetic energy of ions, (B) background from setup components (e.g., the $^{233}U$ source and electronics), (C) background originating from thorium electron shells, and (D) background caused by other radioactive nuclides (descendant nuclides of $^{229}Th$ and $^{232}U$ in this work).

To exclude backgrounds A and B, we changed the ion kinetic energy by varying the MCP surface voltage and obtained MCP counts per ion for $^{229m,g}Th^{2+}$, $^{233}U^{2+}$, $^{229m,g}Th^{+}$, and $^{233}U^{+}$ (Figs. 2c and 2d). Here, the intensities of ion-impact signals at the ion arrival times increased with increasing the ion kinetic energy (decreasing the MCP surface voltage). The intensities of the long-lived decay components observed for $^{229m,g}Th^{2+}$ and $^{229m,g}Th^{+}$ remains constant, independent of the ion kinetic energy, indicating the background A can be excluded[6,7]. Non-observation of the long-lived decay components for $^{233}U^{2+}$ and $^{233}U^{+}$ indicates that the potential backgrounds caused by setup components (background B) can be excluded[6,7].

The potential background by thorium electron shells (background C) for $^{229m,g}Th^{+}$ was excluded by performing measurements using $^{232}Th^{+}$ ions (Methods); the time trace for $^{232}Th^{+}$ at an MCP voltage of −35 V (Extended Data Figs. 1c and 2) had no long-lived decay components.

To discuss the background by the descendant nuclides (background D), we performed measurements of the long-lived decay signals of $^{229m,g}Th^{3+}$ (**Extended Data Fig. 1**a). Here, the molecular ions of descendant nuclides such as $^{213}Bi^{16}O$ hardly form triply charged states due to their high third ionization energies[6], and thus the signals observed for $^{229m,g}Th^{3+}$ ($m/z = 76.3$) do not originate from triply charged molecules of descendant nuclides, but from the IC-electrons of $^{229m}Th$[6,7]. We found that the intensity and lifetime of the long-lived decay component for $^{229m,g}Th^{3+}$ were consistent with those of $^{229m}Th^{2+}$ and $^{229m}Th^{+}$ (**Extended Data Fig. 2**), which excludes background D for $^{229m}Th^{2+}$ and $^{229m}Th^{+}$.

Furthermore, the lifetimes of the long-lived decay components for $^{229m}Th^{2+}$ and $^{229m}Th^{+}$ (Figs. 2a and 2b; Methods) were consistent with the previously reported IC-decay half-life[7]. Consequently, the potential backgrounds were excluded and the long-lived decay components observed for $^{229m,g}Th^{2+}$ and $^{229m,g}Th^{+}$ were consistent with those reported in the previous IC-electron measurements[6,7]. Therefore, we observed IC electrons emitted following the extraction of $^{229m}Th^{+}$ onto an MCP detector.

We note that we performed measurements for $^{229m,g}Th^{+}$ ions directly extracted from the RF-carpet gas cell (Methods) and observed no long-lived decay components (Extended Figs. 1b and 2), which was consistent with the previous measurements[6,7].

## Half-life measurement of $^{229m}Th^{+}$

The observation of the IC electrons following the extraction of $^{229m}Th^{+}$ onto the MCP detector allowed us to count the number of $^{229m}Th^{+}$ ions inside the ion trap. As shown in Fig. 3b, the number of $^{229m}Th^{+}$ in the trap increases over time due to the charge-exchange reaction between $^{229m}Th^{2+}$ and NO gas, while it decreases due to the nuclear decay of $^{229m}Th^{+}$ and chemical reactions with NO and impurity gases. To distinguish the nuclear decay from chemical reactions, we performed comparative measurements for $^{229m,g}Th^{+}$ (predominantly $^{229g}Th^{+}$), where the chemical reaction process (Fig. 3a) is the same as that for $^{229m}Th^{+}$ (Fig. 3b).

We extracted ion bunches of $^{229m,g}Th^{+}$ (bunch width ~0.3 ms) onto the MCP detector after an ion loading time of 0.5 s and a trapping time of 0–16 s. Here, we prevented the trapping of $^{229m,g}Th^{3+}$ ions (Methods), which could be converted to $^{229m,g}Th^{2+}$ and influence the determination of reaction rates.

First, we measured the MCP counts per ion bunch at an MCP voltage of −2000 V for $^{229m,g}Th^+$ (Fig. 3d). The fraction of $^{229m}Th^+$ (2%) in $^{229m,g}Th^+$ was sufficiently small, compared with the ion-source stability (~5%) and statistical uncertainties (several percent), and thus we regarded the ion-impact counts of $^{229m,g}Th^+$ as those of $^{229g}Th^+$. In Fig. 3d, the growth and decay of the number of $^{229g}Th^+$ were observed due to the reactions shown in Fig. 3a. By fitting a function (Eq. (2) in Methods) to the data, we obtained the reaction rates for $^{229g}Th^{2+}$ and $^{229g}Th^+$ as $\mathit{\Gamma}_{2+} = 0.44(6)\ \mathrm{s}^{-1}$ and $\mathit{\Gamma}_{1+} = 0.26(2)\ \mathrm{s}^{-1}$, respectively. Next, Fig. 3f shows the IC electron counts per ion bunch (blue squares) as a function of the trapping time for $^{229m}Th^+$, obtained by subtracting the calculated ion-impact counts per ion bunch (black curve in Fig. 3f; Methods) from the MCP counts per ion bunch measured at −35 V. The shape of the growth and decay curve for $^{229m}Th^+$ (Fig. 3f) is clearly different from that for $^{229g}Th^+$ (Fig. 3d). By fitting a function (Eq. (4) in Methods) to the data in Fig. 3f, we obtained $\mathit{\Gamma}_{2+} + \lambda_{\mathrm{m},2+} = 0.40(5)\ \mathrm{s}^{-1}$ and $\mathit{\Gamma}_{1+} + \lambda_{\mathrm{m},1+} = 1.8(4)\ \mathrm{s}^{-1}$, where $\lambda_{\mathrm{m},2+}$ and $\lambda_{\mathrm{m},1+}$ represent the nuclear decay constants of $^{229m}Th^{2+}$ and $^{229m}Th^+$, respectively.

The values $\mathit{\Gamma}_{2+} = 0.44(6)\ \mathrm{s}^{-1}$ and $\mathit{\Gamma}_{2+} + \lambda_{\mathrm{m},2+} = 0.40(5)\ \mathrm{s}^{-1}$ are consistent within the uncertainties, indicating that $\lambda_{\mathrm{m},2+}$ is negligibly small. This agrees with the previous reports[6,13] on the half-life of $^{229m}Th^{2+}$ ($\lambda_{\mathrm{m},2+} < 0.012\ \mathrm{s}^{-1}$). We independently measured the variation in the number of $^{229m,g}Th^{2+}$ and $^{229m}Th^{2+}$ over the trapping time (Figs. 3c and 3e) and obtained $\mathit{\Gamma}_{2+} = 0.425(6)\ \mathrm{s}^{-1}$ and $\mathit{\Gamma}_{2+} + \lambda_{\mathrm{m},2+} = 0.45(3)\ \mathrm{s}^{-1}$, which were consistent with those obtained from the measurements of $^{229m,g}Th^+$.

Subtracting $\mathit{\Gamma}_{1+} = 0.26(2)\ \mathrm{s}^{-1}$ from $\mathit{\Gamma}_{1+} + \lambda_{\mathrm{m},1+} = 1.8(4)\ \mathrm{s}^{-1}$ yielded $\lambda_{\mathrm{m},1+} = 1.5(4)\ \mathrm{s}^{-1}$. We also performed measurements at different NO-gas pressures (Extended Data Tables 1 and 2), and the obtained $\lambda_{\mathrm{m},1+}$ values were consistent within the uncertainties. Taking the weighted average of these values and accounting for the potential systematic error of $0.17\ \mathrm{s}^{-1}$ (Methods), we determined the nuclear decay constant of $^{229m}Th^+$ to be $\lambda_{\mathrm{m},1+} = 1.50(26)\ \mathrm{s}^{-1}$ at $p_{\mathrm{total}} = 0.40$ Pa , corresponding to a half-life of $T_{1/2} = 0.46(8)$ s. Furthermore, we performed measurements at different Ar-gas pressures (Extended Data Fig. 3) and obtained $\lambda_{\mathrm{m,1+}} = 1.9(15)$ and $1.8(6)\ \mathrm{s}^{-1}$ at $p_{\mathrm{total}}$ = 0.20 and 0.50 Pa, respectively, which were consistent with the value at $p_{\mathrm{total}}$ = 0.40 Pa.

**Discussion of the $^{229\mathrm{m}}Th^{+}$ half-life and future prospects**

The half-life of 0.46(8) s at $p_{\mathrm{total}}$ = 0.40 Pa for $^{229\mathrm{m}}Th^{+}$ is $7\times10^4$ times longer than that of the IC decay (7(1) μs[7]) and ~$10^3$ times shorter than that of the radiative decay (100–670 s for solid-state $^{229\mathrm{m}}Th$ [2–4,8,9,11] and $1400^{+600}_{-300}$ s for $^{229\mathrm{m}}Th^{3+}$ [12]). In our experiment, the IC decay was prohibited because the atomic electrons in $^{229\mathrm{m}}Th^{+}$ were cooled to the electronic ground state (admixture of [Rn]$6d^2 7s$, [Rn]$6d7s^2$, and others[24]; total angular momentum $J$ = 3/2; ionization energy 12.1(2) eV[27]) by rapid collisional quenching with Ar gas[35]. It was reported that the collision between $Th^{+}$ and Ar slightly populated the 0.19-eV electronic excited state ($J$ = 5/2)[35], although the IC decay remains prohibited in this case. The prohibition of the IC decay and the significant difference in half-life compared with the radiative decay suggest that $^{229\mathrm{m}}Th^{+}$ dominantly deexcites through a new decay mode, probably the EB process, as theoretically predicted[20,24–26]. We should note that when the nuclear transition energy matches an electronic transition energy within their natural linewidths, the bound internal conversion process that does not involve the emission of a photon[36] could occur; however, based on laser spectroscopy of $^{232}Th^{+}$ [28,35], the probability of such energy matching can be excluded. Furthermore, it is difficult to consider that the transition energy matching caused by collisional broadening of electronic states[25] occurred, because the collisional broadening is expected to be comparable to or smaller than the natural linewidth of electronic states in $^{229\mathrm{m}}Th^{+}$ under our experimental conditions[35].

To compare with theoretical calculations, we introduce the ratio ($R$) of the decay constant of $^{229\mathrm{m}}Th^{+}$ ($\lambda_{\mathrm{m,1+}}$) to the radiative decay constant ($\lambda_\gamma$), namely, $R = \lambda_{\mathrm{m,1+}}/\lambda_\gamma$. If we use the decay constant of $^{229\mathrm{m}}Th^{3+}$ $(5.1(15) \times 10^{-4}\ \mathrm{s}^{-1})$[12] as $\lambda_\gamma$, the $R$ value for $^{229\mathrm{m}}Th^{+}$ is calculated to be $R = 2.9(10) \times 10^3$. Recent theoretical studies[25,26] calculated the EB-decay rate of $^{229\mathrm{m}}Th^{+}$ at an isomer energy of 8.3557 eV[2,3] and reported $R = 5.7 \times 10^4$ (Ref. [25]) and $R = 2.660 \times 10^2 - 1.310 \times 10^5$ (Ref. [26]), which are not far from our experimental $R$ value. The large range of the theoretical $R$ values likely originate from the difficulty in accurately calculating the electronic excitation energies of $Th^+$ as well as the ambiguity of the angular momenta of the excited states[26]. Further theoretical studies based on our experimental value will lead to a better understanding of the EB decay of $^{229m}Th^+$.

To directly confirm the existence of the EB decay and to understand the decay mechanism, we plan to transport mass-separated $^{229m}Th^+$ ions to another ion trap, where we will detect photons emitted in the EB decay and analyze their wavelengths. Studies on EB decay are also useful for practical applications such as the $^{229}Th$ nuclear clock. In the solid-state nuclear clock, after the isomeric state of $^{229}Th$ with a long radiative half-life is populated by a vacuum ultraviolet laser, it is desirable to return the isomer to the ground state as quickly as possible for the next interrogation of the resonant laser frequency. The decay of $^{229m}Th$ may be accelerated if we actively induce EB decay by controlling the electronic states of $^{229m}Th$. For example, a recent study reported the acceleration of the decay of $^{229m}Th$ embedded in $CaF_2$ crystals by increasing the flux of X-ray beams hitting the crystals[9], which may be related to the EB decay. Furthermore, understanding the EB process will enable the inverse EB process[23,26,37,38], which may allow excitation to the isomeric state more efficiently than direct laser excitation. A large number of $^{229m}Th$ produced by the inverse EB process should be useful for future $^{229m}Th$ studies, such as investigating the chemical effects on nuclear decay and determining nuclear properties more precisely (e.g., radiative half-life, nuclear charge radius, and quadrupole moment[12,13]).

**Figures**

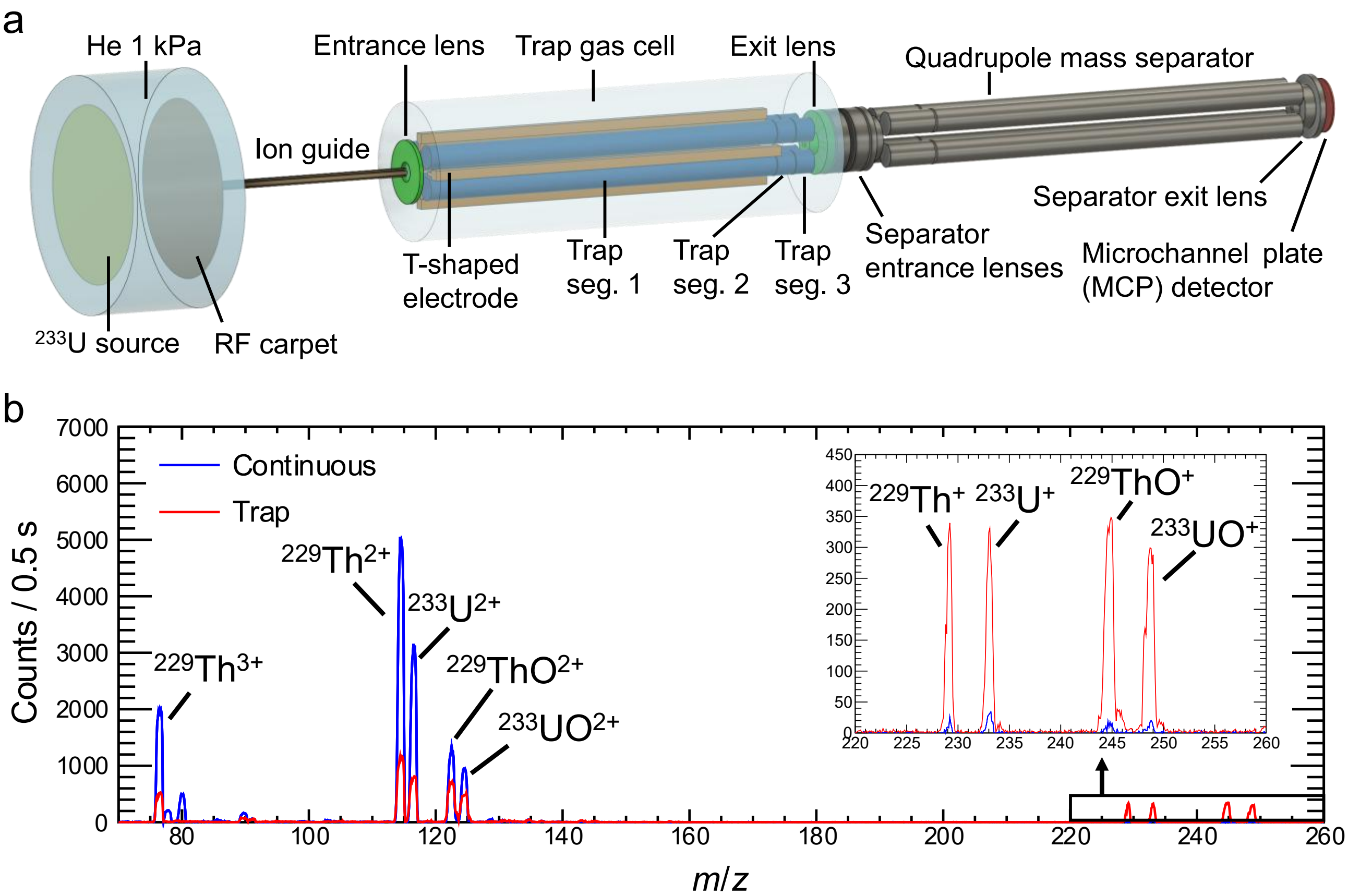

**Fig. 1. Extraction, trapping, and mass separation of $^{229}$Th ions. a**, Schematic of the experimental setup. $^{229}$Th ions recoiling from a $^{233}$U source were extracted with the RF-carpet gas cell and the quadrupole ion guide into the ion trap. The ions extracted from the ion trap were mass-separated with the quadrupole mass separator and transported to the microchannel plate (MCP) detector. **b**, Mass-to-charge ratio spectra for continuously extracted ions (blue) and ions extracted after 0.5-s loading and 0.1-s trapping (red). The inset shows the spectrum in the range of *m*/*z* = 220–260 on an expanded scale. The trap gas cell was filled with a mixture of He, Ar, and NO gases at pressures of ~0.16, ~0.24, and 6.1(3)×10$^{-6}$ Pa, respectively ($p_{\mathrm{total}}$ = 0.40 Pa).

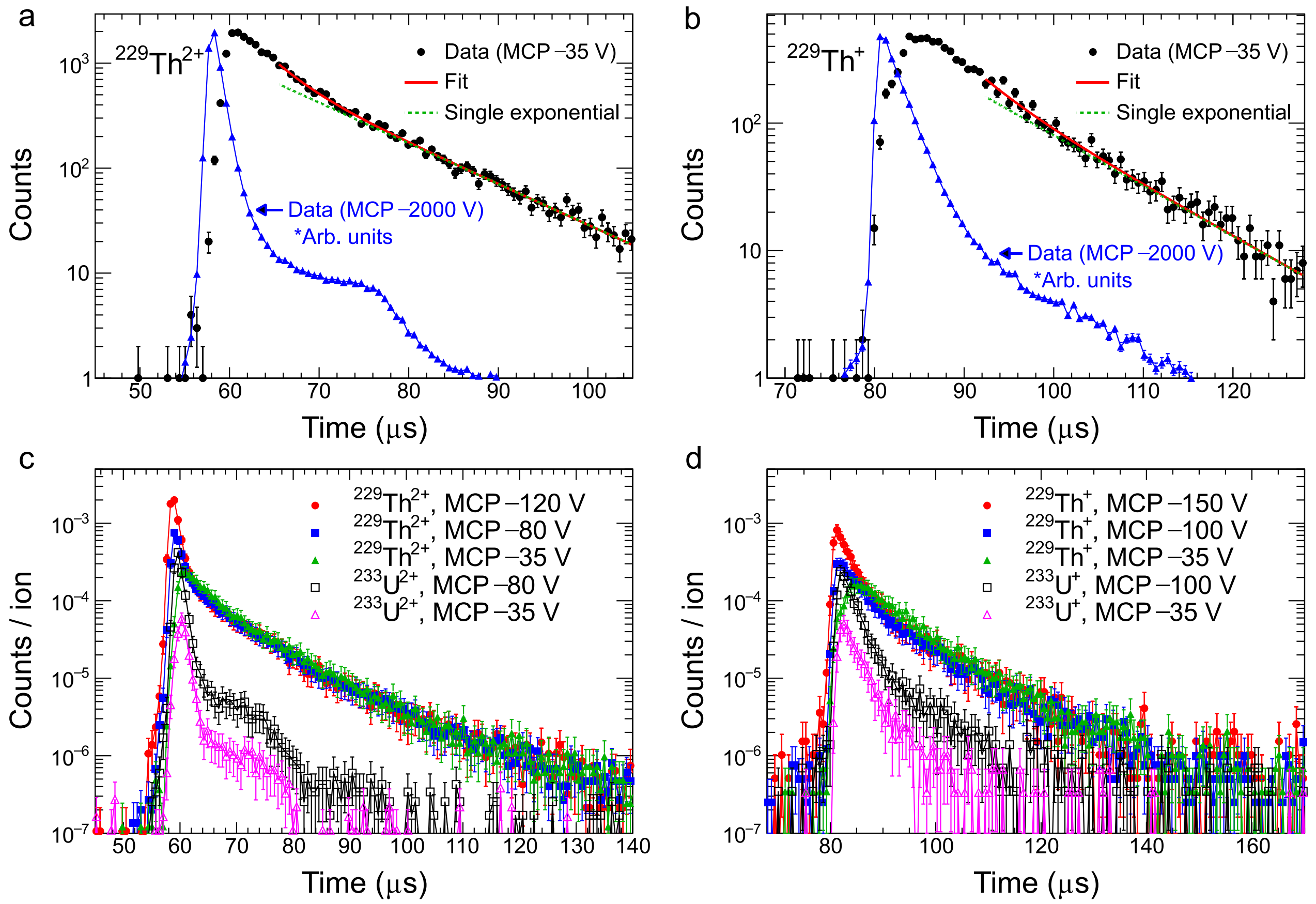

**Fig. 2. Time traces of the MCP signals for the extraction of short ion bunches. a–b**, MCP counts (black circle) as a function of time (bin width 655 ns) at an MCP surface voltage of −35 V for $^{229m,g}Th^{2+}$ (a) and $^{229m,g}Th^{+}$ (b). Double exponential decay functions (red solid curve) were fitted to the data. Single exponential components with longer half-lives in the fitted functions are shown as green dashed curves. The blue triangles show the time traces obtained at an MCP surface voltage of −2000 V in arbitrary units. **c**, MCP counts divided by the total number of ions extracted on the MCP detector as a function of time (bin width 655 ns) for $^{229m,g}Th^{2+}$ at MCP surface voltages of −120 (red circle), −80 (blue closed square), and −35 V (green closed triangle), and those for $^{233}U^{2+}$ at −80 (black open square) and −35 V (purple open triangle). **d**, MCP counts divided by the total number of ions extracted on the MCP detector as a function of time (bin width 655 ns) for $^{229m,g}Th^{+}$ at MCP surface voltages of −150 (red circle), −100 (blue closed square), and −35 V (green closed triangle), and those for $^{233}U^{+}$ at −100 (black open square) and −35 V (purple open triangle). The total number of ions in (c)–(d) was obtained by multiplying the number of ions per bunch, measured at −2000 V, by the total number of extracted bunches. The 0 μs in (a)–(d) corresponds to the time when ion extraction from the ion trap was initiated. The error bars in (a)–(d) represent the 1 s.d. statistical uncertainties. The gas pressures in the ion trap were kept at $p_{total}$ = 0.40 Pa, $p_{He}$ ≈ 0.16 Pa, $p_{Ar}$ ≈ 0.24 Pa, and $p_{NO}$ = 6.1(3)×$10^{-6}$ Pa.

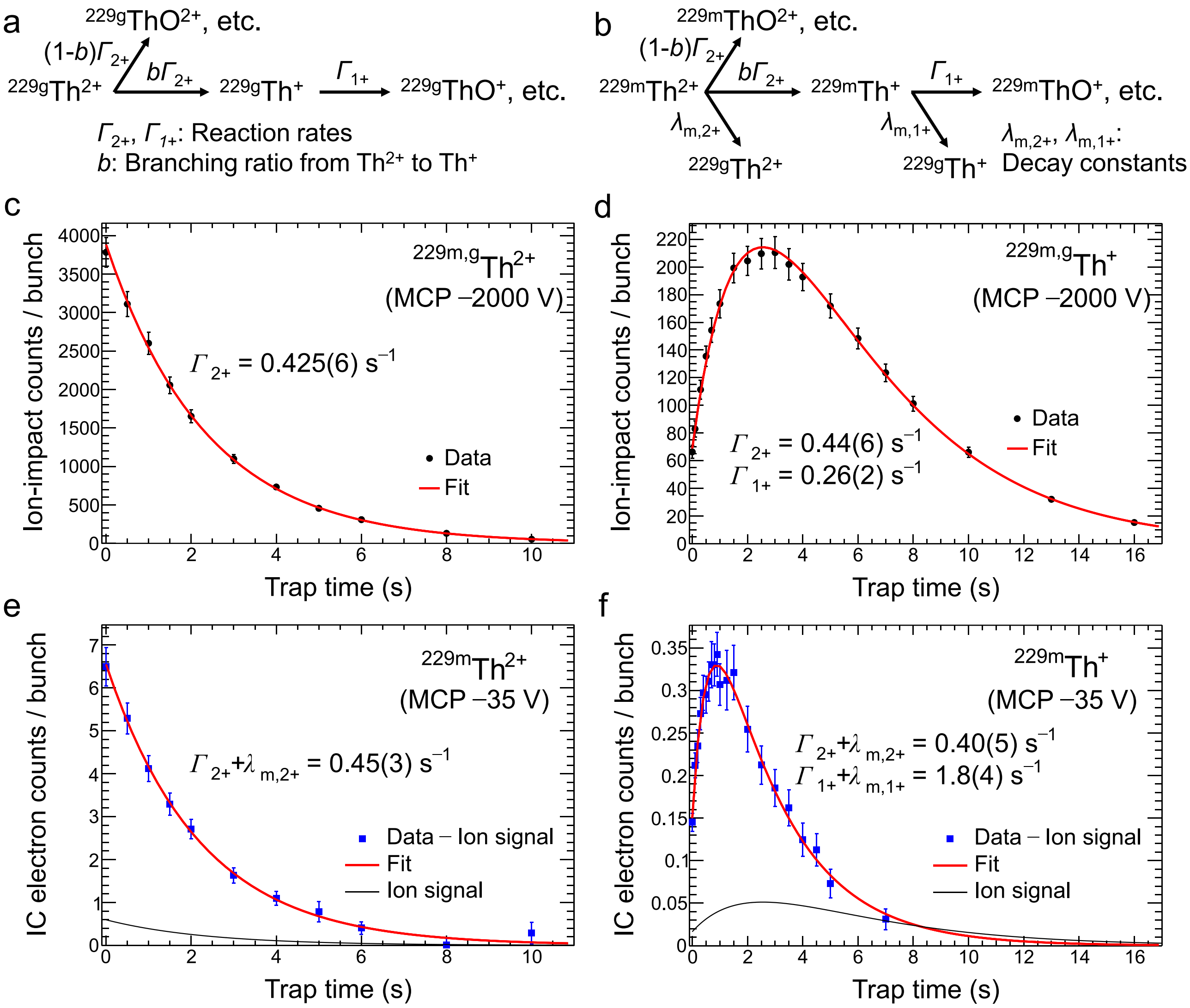

**Fig. 3. Measurement of the number of $^{229m,g}Th^{2+}$ and $^{229m,g}Th^{+}$ ions as a function of the trapping time. a**, Reaction scheme starting from $^{229g}Th^{2+}$. $^{229g}Th^{2+}$ is converted to other species with a reaction rate $\Gamma_{2+}$, and $^{229g}Th^{+}$ is produced from $^{229g}Th^{2+}$ with a branching ratio $b$. $^{229g}Th^{+}$ is converted to $^{229g}ThO^{+}$ or other species with a reaction rate $\Gamma_{1+}$. **b**, Reaction scheme starting from $^{229m}Th^{2+}$ and the nuclear decay of $^{229m}Th^{2+}$ and $^{229m}Th^{+}$. The nuclear decay constants for $^{229m}Th^{2+}$ and $^{229m}Th^{+}$ are expressed as $\lambda_{m,2+}$ and $\lambda_{m,1+}$, respectively. **c–d**, Ion-impact counts (MCP counts at −2000 V) per ion bunch (black circle) as a function of the trapping time for $^{229m,g}Th^{2+}$ (c) and $^{229m,g}Th^{+}$ (d), and the functions (red curve) obtained by fitting Eqs. (1) and (2), respectively, to the data. Here, the ion-impact counts of $^{229m,g}Th^{2+}$ and $^{229m,g}Th^{+}$ were regarded as those of $^{229g}Th^{2+}$ and $^{229g}Th^{+}$, respectively, due to the negligibly small fraction of $^{229m}Th$ (~2%) in $^{229m,g}Th$. **e–f**, IC-

electron counts per ion bunch (blue square) as a function of the trapping time for $^{229m}Th^{2+}$ (e) and $^{229m}Th^{+}$ (f), and the functions (red curve) obtained by fitting Eqs. (3) and (4), respectively, to the data. IC-electron counts were obtained by subtracting the ion-impact counts (black curve) from the MCP counts at an MCP surface voltage of −35 V. The error bars in (c)–(f) include 1 s.d. statistical uncertainties and fluctuations in the number of $^{229}$Th ions supplied by our RF carpet gas cell (5%). The gas pressures in the ion trap were kept at $p_{total} = 0.40$ Pa, $p_{He} \approx 0.16$ Pa, $p_{Ar} \approx 0.24$ Pa, and $p_{NO} = 9.5(5)\times10^{-7}$ Pa.

## Methods

### Experimental setup

We used a $^{233}U$ source (600 kBq, diameter 90 mm) electrodeposited on a titanium foil (thickness 50 μm)[12] to produce $^{229m,g}Th$ ions. The amount of $^{232}U$ contained in the source was 0.226(2) ppm. The chemical purification and electrodeposition of $^{233}U$ were performed 34 months before the main measurements in this study (Campaign A). Additional experiments at $p_{total}$ = 0.20 and 0.50 Pa (Campaign B) and the measurements using $^{229m,g}Th^{3+}$ and $^{229m,g}Th^{+}$ directly extracted from the RF-carpet gas cell (Campaign C) were performed 52 months after the chemical purification of $^{233}U$. We also performed measurements using $^{232}Th^{+}$ (Campaign D).

In Campaign A, the $^{233}U$ source was placed in the gas cell, where an RF carpet[12,32,39] (diameter 90 mm) was mounted (**Fig. 1**a). The distance between the $^{233}U$ source and the RF carpet was 5 cm. We applied DC voltages of +40 and +2 V to the $^{233}U$ source and the RF carpet, respectively, creating a DC gradient of 7.6 V/cm. The $^{229m,g}Th$ ions emitted from the $^{233}U$ source by the recoil energy of α decay were thermalized in He gas at 1 kPa and pushed to the RF carpet by the DC gradient. The RF voltages applied to the RF carpet (9.61 MHz, 220 $V_{pp}$ on average) prevented the ions from colliding with the electrodes of the RF carpet. To transport the ions to the central hole of the RF carpet, we used the RF-surfing technique[12,33,34,39]; a traveling wave potential was created by applying phase-shifted audio frequency voltages (200 kHz, 10.3 $V_{pp}$) to the RF carpet. Just before Campaigns B and C, we relocated the apparatus, replaced several components such as a He-gas purifier, and modified some experimental parameters; here we confirmed that the extraction rates of $^{229m,g}Th^{2+}$ were similar to those in Campaign A. In Campaign C, a large number of $^{229m,g}Th^{+}$ (~200 ions/s), compared with Campaigns A and B, could be extracted from the RF-carpet gas cell by adjusting the amplitudes of the audio frequency voltages. In Campaign D, the $^{233}U$ source was replaced with a $^{232}Th$ metallic foil, and $^{232}Th^{+}$ ions were produced by laser ablation of $^{232}Th$ using a nitrogen laser and extracted from the RF-carpet gas cell.

The ions extracted from the central hole of the RF carpet were transported by the quadrupole ion guide[39] (length 123 mm), where RF voltages of 130 $V_{pp}$ (2.8 MHz) were applied. The electrodes of the ion guide were segmented to create a DC gradient; in this study, the DC voltages of +0.97 and 0 V were applied to the first (upstream) and last (downstream) segments, respectively. The ions extracted from the ion guide were sent to the quadrupole ion trap with three segments (**Fig. 1**a). The diameter of the quadrupoles was 12 mm, and the diameter of the trap area was 10.5 mm. The lengths of segments 1, 2, and 3 were 192, 10, and 10 mm, respectively. We applied RF voltages of ~510 or ~900 $V_{pp}$ with a frequency of 1.38 MHz to the trap quadrupoles. In segment 1, T-shaped electrodes were installed to make a DC gradient, which rapidly transported the ions downstream in the presence of the gas. We placed the trap exit lens downstream of trap segment 3, where a DC voltage of −35 V was applied for extracting the ions from the ion trap, while +1 or +15 V was applied for trapping the ions.

We introduced a mixture of Ar and NO gases into the ion-trap region through a variable leak valve. A buffer tank was installed upstream of the variable leak valve to mix the Ar and NO gases. We introduced Ar gas (purity >99.9999%, Taiyo Nippon Sanso) into the buffer tank through a mass flow controller. We also introduced NO gas (purity >99%, Taiyo Nippon Sanso), which mainly contained $NO_2$ and $N_2$ gases as impurities, into the buffer tank through a variable leak valve. The buffer tank was always evacuated with a dry pump through a piezo valve; the pressure of the buffer tank was controlled so that the total pressure of the ion-trap region was kept at $p_{\mathrm{total}}$ = 0.40 Pa in Campaign A. The concentration of the NO gas in the buffer tank was changed by varying the flow rate of the Ar gas ($f_{\mathrm{Ar}}$) introduced into the buffer tank in the range of 5–30 mL/min. We estimated the partial pressure of the NO gas ($p_{\mathrm{NO}}$) in the ion trap region by measuring the loss rate of the $^{229\mathrm{m,g}}Th^{2+}$ ions, which is expressed as $\Gamma_{2+} = k_{2+,\mathrm{NO}} n_{\mathrm{NO}} + \sum_{\mathrm{i}} k_{2+,\mathrm{i}} n_{\mathrm{i}}$. Here, $k_{2+,\mathrm{NO}}$ is the reaction rate constant between $^{229\mathrm{m,g}}Th^{2+}$ and NO, $n_{\mathrm{NO}}$ is the density of the NO gas, $k_{2+,\mathrm{i}}$ is the reaction rate constant between $^{229\mathrm{m,g}}Th^{2+}$ and an impurity gas, and $n_{\mathrm{i}}$ is the density of the impurity gas. We found that the $\Gamma_{2+}$ values were proportional to $1/f_{\mathrm{Ar}}$; here, we obtained $\sum_{\mathrm{i}} k_{2+,\mathrm{i}} n_{\mathrm{i}} = 0.29(1)\ \mathrm{s}^{-1}$ and $k_{2+,\mathrm{NO}} n_{\mathrm{NO}} = 3.7(2)/f_{\mathrm{Ar}}$.

In a previous study[31], the $k_{2+,\mathrm{NO}}$ value was measured to be 5.3×10$^{-9}$ cm$^3$ molecule$^{-1}$ s$^{-1}$. From this $k_{2+,\mathrm{NO}}$ value and the Ar flow rate ($f_{\mathrm{Ar}}$), we obtained the $n_{\mathrm{NO}}$ value, which was converted to the $p_{\mathrm{NO}}$ value. Since the $k_{2+,\mathrm{NO}}$ value measured in the previous study[31] has an error of 50%, the $p_{\mathrm{NO}}$ values obtained in this study have systematic errors (50%). We should note that the precision of the $p_{\mathrm{NO}}$ values does not affect the result of the $\lambda_{\mathrm{m},1+}$ values measured in this study. In Campaign B, we set $p_{\mathrm{total}}$ = 0.20 or 0.50 Pa, where $p_{\mathrm{Ar}} \approx$ 0.08 or 0.38 Pa, $p_{\mathrm{NO}} \approx$ 1.8(3)×10$^{-7}$ or 4.6(6)×10$^{-7}$ Pa, respectively, and $\sum_{\mathrm{i}} k_{2+,\mathrm{i}} n_{\mathrm{i}} = 0.053(6)\ \mathrm{s}^{-1}$. In Campaigns C and D, we did not introduce NO gas and maintained the total pressure at 0.40 Pa ($p_{\mathrm{Ar}} \approx$ 0.28 Pa).

The ions extracted from the ion trap were sent to the quadrupole mass separator (Ardara Technologies), equipped with entrance lenses, prefilters, and the exit lens (**Fig. 1**a). The mass resolutions for $^{229\mathrm{m,g}}\mathrm{Th}^{2+}$ and $^{229\mathrm{m,g}}\mathrm{Th}^{+}$ were ~1 amu and ~0.7 amu in the full width at half maximum, respectively (**Fig. 1**b). The ions were finally transported to the MCP detector (F9890-31, Hamamatsu Photonics; effective area diameter 27 mm). We applied −2000 V to the surface of the MCP detector for ion detection, while we applied −35 V for the IC-electron detection of $^{229\mathrm{m}}\mathrm{Th}$. For all the measurements in this study, the potential difference between the surface (input) and the output of the MCP detector was set to 2000 V, and that between the output and the anode was set to 500 V. The signals from the anode of the MCP detector were sent to a multichannel scaler (MCS8A, FAST ComTec) for counting the signals.

### Preparation of short ion bunches and IC-electron measurement

We prepared short ion bunches of $^{229\mathrm{m,g}}\mathrm{Th}^{2+}$ and $^{229\mathrm{m,g}}\mathrm{Th}^{+}$ to detect the IC electrons of $^{229\mathrm{m}}\mathrm{Th}$ in Campaign A as follows. First, ions from the RF-carpet gas cell were loaded into the ion trap at gas pressures of $p_{\mathrm{total}}$ = 0.40 Pa, $p_{\mathrm{He}} \approx$ 0.16 Pa, $p_{\mathrm{Ar}} \approx$ 0.24 Pa, and $p_{\mathrm{NO}}$ = 6.1(3)×10$^{-6}$ Pa. The ion loading times were 0.1 and 0.7 s for the $^{229\mathrm{m,g}}\mathrm{Th}^{2+}$ and $^{229\mathrm{m,g}}\mathrm{Th}^{+}$ measurements, respectively. The longer loading time for $^{229m,g}Th^{+}$ (0.7 s) was to increase the number of $^{229m,g}Th^{+}$ ions in the trap by the charge-exchange reaction. The amplitude of the RF voltages applied to the trap quadrupoles was set to ~510 $V_{pp}$ (1.38 MHz). After stopping the ion loading, the ions were trapped for 0.1 s. During the ion loading and trapping, the ions were pushed toward the downstream of trap segment 1 by a DC gradient (~0.4 V/cm) created by the T-shaped electrodes (**Fig. 1**a). The ions accumulated in the trap were extracted by switching the DC voltages applied to the exit lens, trap segment 1, and the T-shaped electrodes from +1, −2, and −20 V to −35, +35, and +35 V, respectively, within a microsecond. The ion bunch was then subjected to mass separation and transported to the MCP detector. The trapping and extraction of the ion bunch were repeated with cycle times of 0.2 and 0.8 s for $^{229m,g}Th^{2+}$ and $^{229m,g}Th^{+}$, respectively. Each ion bunch contained ~320 ions of $^{229m,g}Th^{2+}$ (~6 ions of $^{229m}Th^{2+}$) and ~200 ions of $^{229m,g}Th^{+}$ (~4 ions of $^{229m}Th^{+}$) as well as ~250 ions of $^{233}U^{2+}$ and ~130 ions of $^{233}U^{+}$. The signals from the MCP detector were recorded in bins of 655 ns with the multichannel scaler. The recording for each ion bunch was started at the time of ion extraction from the ion trap. The results of the measurements are shown in **Fig. 2**.

In similar ways, we prepared short ion bunches for $^{229m,g}Th^{3+}$ and $^{229m,g}Th^{+}$ extracted from the RF-carpet gas cell (Campaign C), as well as for $^{232}Th^{+}$ ions (Campaign D). The results of the measurements at MCP voltages of −2000 and −35 V are shown in **Extended Data Fig. 1**.

We also performed measurements using short ion bunches of $^{229m,g}Th^{2+}$ in Campaign C (52 months from $^{233}U$ purification), and the intensity of the long-lived decay component was consistent with that in Campaign A (34 months from $^{233}U$ purification). If the long-lived decay components had originated from descendant nuclides of $^{229}Th$ and $^{232}U$, the intensity of the component in Campaign C should have become 1.4–1.5 times higher than that in Campaign A due to the growth of the descendant nuclides after chemical purification of $^{233}U$. Hence, background D can be excluded.

We obtained the lifetimes of the decay components observed for $^{229m,g}Th^{2+}$ and $^{229m,g}Th^{+}$ by fitting exponential decay functions to the data in Figs. 2a and 2b. Here, the left boundaries of the fitting ranges were selected such that the ion-impact signals measured at −2000 V were negligibly small. Double exponential decay functions provided better fits than single exponential decay functions, and we obtained half-lives of 7.8(3) and 2.3(5) μs for $^{229m,g}Th^{2+}$ (Fig. 2a). The half-life of 7.8(3) μs agrees with the previously reported IC-decay half-life of 7(1) μs[7]. For $^{229m,g}Th^{+}$ (Fig. 2b), we obtained half-lives of 7.6(7) and 2.9(14) μs, consistent with the values for $^{229m,g}Th^{2+}$. The half-life of 7.6(7) for $^{229m,g}Th^{+}$ is also consistent with the IC-decay half-life, supporting the interpretation that the long-lived decay component originates from the IC electrons of $^{229m}Th$. We note that the short-lived decay component for $^{229m,g}Th^{2+}$ (half-life: 2.3(5) μs) seemed to be observed in the previous study (Figs. 2 and 4 in Ref. [7]). Since no such component was observed for the extraction of $^{230}Th^{2+}$ ions[7], the short-lived component should be attributed to the decay of $^{229m}Th$, although further studies are required to clarify its origin.

**Measurement of MCP counts as a function of the trapping time**

In Campaign A, we measured the MCP counts as a function of the trapping time for $^{229m,g}Th^{2+}$ and $^{229m,g}Th^{+}$ at MCP surface voltages of −2000 and −35 V at $p_{total}$ = 0.40 Pa (Figs. 3c–3f). In this experiment, ions from the RF-carpet gas cell were loaded into trap segment 1 for 0.5 s. The ions were then trapped in the range of 0–16 s after we stopped the ion loading. During the ion loading and trapping, the DC potential of trap segment 1 was set to −50 V, which was much lower than the DC potentials of the entrance lens (−6 V for loading and +5 V for trapping), trap segments 2 and 3 (same potential; −2 or −7 V), and the exit lens (+15 V). The amplitude of the RF voltages applied to the trap quadrupoles was set to ~900 $V_{pp}$ (1.38 MHz), which was high enough to exclude $^{229m,g}Th^{3+}$ from the ion trap. After the ion loading and trapping, the ions were extracted from the trap by switching the DC voltages of trap segment 1 and the exit lens to −1 and −35 V, respectively. We confirmed that all the ions trapped in trap segment 1 were extracted within 0.5 s owing to the DC gradient created by the T-shaped electrodes. The ions extracted from the trap were mass-separated, and the $^{229m,g}Th^{2+}$ or $^{229m,g}Th^{+}$ ions were transported to the MCP detector whose surface voltage was set to −2000 or −35 V. We performed this measurement at different NO gas pressures, namely, $p_{NO}$ = 9.5(5)×10$^{-7}$, 1.15(6)×10$^{-6}$, 1.44(7)×10$^{-6}$, and 2.94(15)×10$^{-6}$ Pa (Extended Data Tables 1 and 2).

As shown in Figs. 3c and 3d, we measured the MCP counts per ion bunch as a function of the trapping time $t$ at an MCP surface voltage of −2000 V. The data for $^{229m,g}Th^{2+}$ (predominantly $^{229g}Th^{2+}$) shown in **Fig. 3**c were fitted with a single exponential decay function, namely,

$$f_{2+,g} = p_{2+,g} e^{-\Gamma_{2+} t}, \qquad (1)$$

where $p_{2+,g}$ and $\Gamma_{2+}$ were free parameters in the fitting. The data for $^{229m,g}Th^{+}$ (predominantly $^{229g}Th^{+}$) shown in **Fig. 3**d were fitted with a function

$$f_{1+,g} = p_{1+,g} \frac{\Gamma_{2+}}{\Gamma_{1+} - \Gamma_{2+}} \left(e^{-\Gamma_{2+} t} - e^{-\Gamma_{1+} t}\right) + p_{1+,g}' e^{-\Gamma_{1+} t}, \qquad (2)$$

where $p_{1+,g}$, $p_{1+,g}'$, $\Gamma_{2+}$, and $\Gamma_{1+}$ were free parameters in the fitting. Note that the branching ratio $b$ from $^{229m,g}Th^{2+}$ to $^{229m,g}Th^{+}$ (Figs. 3a and 3b) is included in the parameter $p_{1+,g}$. Consequently, we obtained the loss rates of $^{229m,g}Th^{2+}$ ($\Gamma_{2+}$) and $^{229m,g}Th^{+}$ ($\Gamma_{1+}$) due to chemical reactions (Figs. 3a and 3b).

The MCP counts at an MCP voltage of −35 V (Figs. 3e and 3f) contained IC-electron signals of $^{229m}Th$ and ion-impact signals. To subtract the ion-impact signals, we introduce a parameter $\epsilon_{ion}$, which is the ratio of the ion-impact MCP counts at −35 V to those at −2000 V. We measured the $\epsilon_{ion}$ values for $^{233}U^{2+}$ and $^{233}U^{+}$ to be 1.5(5)×10$^{-4}$ and 2.5(3)×10$^{-4}$, respectively. Moreover, in Campaign D, we measured the $\epsilon_{ion}$ value for $^{232}Th^{+}$ to be 2.4(5)×10$^{-4}$, which was consistent with that for $^{233}U^{+}$. This indicates that the measured $\epsilon_{ion}$ values for $^{233}U$ and $^{232}Th$ do not contain signals originating from electron shells of uranium and thorium but only the ion-impact signals. We used the $\epsilon_{ion}$ values of 1.5(5)×10$^{-4}$ (from $^{233}U^{2+}$) and 2.4(5)×10$^{-4}$ (from $^{232}Th^{+}$) for analyzing the data of $^{229m,g}Th^{2+}$ and $^{229m,g}Th^{+}$, respectively. By multiplying the fitted functions in Figs. 3c and 3d by $\epsilon_{ion}$, we obtained the ion-impact counts as a function of the trapping time at −35 V, which are shown as black curves in Figs.

3e and 3f. Here, we took into account the error propagation arising from the uncertainties in the fitted functions and $\epsilon_{\mathrm{ion}}$. Consequently, we subtracted the calculated ion-impact counts from the MCP counts at −35 V and obtained the IC-electron counts shown as blue squares in Figs. 3e and 3f. The IC-electron counts (the number of trapped $^{229\mathrm{m}}$Th ions) as a function of the trapping time for $^{229\mathrm{m}}\mathrm{Th}^{2+}$ (**Fig. 3**e) were fitted with a function

$$f_{2+,\mathrm{m}} = p_{2+,\mathrm{m}} e^{-(\Gamma_{2+}+\lambda_{\mathrm{m},2+})t}, \qquad (3)$$

where $p_{2+,\mathrm{m}}$ and $\Gamma_{2+}+\lambda_{\mathrm{m},2+}$ were free parameters in the fitting. The IC-electron counts as a function of the trapping time for $^{229\mathrm{m}}\mathrm{Th}^{+}$ (**Fig. 3**f) were fitted with a function

$$f_{1+,\mathrm{m}} = p_{1+,\mathrm{m}} \frac{\Gamma_{2+}+\lambda_{\mathrm{m},2+}}{(\Gamma_{1+}+\lambda_{\mathrm{m},1+})-(\Gamma_{2+}+\lambda_{\mathrm{m},2+})}\left(e^{-(\Gamma_{2+}+\lambda_{\mathrm{m},2+})t} - e^{-(\Gamma_{1+}+\lambda_{\mathrm{m},1+})t}\right) + p_{1+,\mathrm{m}}' e^{-(\Gamma_{1+}+\lambda_{\mathrm{m},1+})t}, \quad (4)$$

where $p_{1+,\mathrm{m}}$, $p_{1+,\mathrm{m}}'$, $\Gamma_{2+}+\lambda_{\mathrm{m},2+}$, and $\Gamma_{1+}+\lambda_{\mathrm{m},1+}$ were free parameters in the fitting. We were able to evaluate the nuclear decay constant $\lambda_{\mathrm{m},1+}$ by subtracting the $\Gamma_{1+}$ value (derived from the data at −2000 V) from the $\Gamma_{1+}+\lambda_{\mathrm{m},1+}$ value (derived from the data at −35 V).

The measurements at different NO-gas pressures gave the results shown in Extended Data Tables 1 and 2 for $^{229\mathrm{m,g}}\mathrm{Th}^{2+}$ and $^{229\mathrm{m,g}}\mathrm{Th}^{+}$, respectively. The $\Gamma_{2+}$ and $\Gamma_{2+}+\lambda_{\mathrm{m},2+}$ values were in agreement at each NO pressure, which is consistent with the long half-life of $^{229\mathrm{m}}\mathrm{Th}^{2+}$ (>60 s[6,13]). The $\lambda_{\mathrm{m},1+}$ values obtained at four NO-gas pressures (**Extended Data Table 2**) were consistent within the uncertainties. The weighted average of these values gave $\lambda_{\mathrm{m},1+} = 1.50(20)\ \mathrm{s}^{-1}$. To evaluate the systematic errors in $\lambda_{\mathrm{m},1+}$, we fitted the Eqs. (2) and (4) to the $^{229\mathrm{m,g}}\mathrm{Th}^{+}$ data obtained at −2000 and −35 V, respectively, while fixing $\Gamma_{2+}$ and $\Gamma_{2+}+\lambda_{\mathrm{m},2+}$ to the $\Gamma_{2+}$ values obtained from the $^{229\mathrm{m,g}}\mathrm{Th}^{2+}$ data at −2000 V (e.g., Fig. 3c). This analysis for all NO-gas pressures yielded a maximum systematic error of 0.17 $\mathrm{s}^{-1}$ in $\lambda_{\mathrm{m},1+}$. Finally, we obtained $\lambda_{\mathrm{m},1+} = 1.50(26)\ \mathrm{s}^{-1}$ as the nuclear decay constant of $^{229\mathrm{m}}\mathrm{Th}^{+}$ at $p_{\mathrm{total}}$ = 0.40 Pa. In Campaign B, we measured the decay constants of $^{229\mathrm{m}}\mathrm{Th}^{+}$ at $p_{\mathrm{total}}$ = 0.20 and 0.50 Pa in similar ways (**Extended Data Fig. 3** shows an example of the data), yielding $\lambda_{\mathrm{m},1+} = 1.9(15)$ and $1.8(6)\ \mathrm{s}^{-1}$, respectively. We note that the measurements in Campaign B were performed under much lower partial pressures of NO and impurity molecules than those in Campaign A. The consistency of the obtained $\lambda_{\mathrm{m,1+}}$ values between Campaigns A and B suggests that the influence of NO, impurity molecules, and the buffer gas on $\lambda_{\mathrm{m,1+}}$ is negligibly small.

Molecular ions of the descendant nuclides of $^{229}$Th and $^{232}$U, such as $^{213}$Bi$^{16}$O$^{2+}$ and $^{213}$Bi$^{16}$O$^{+}$, could be produced during ion trapping and potentially interfere with the data of $^{229\mathrm{m,g}}$Th$^{2+}$ and $^{229\mathrm{m,g}}$Th$^{+}$ (Figs. 3c–3f); however, we consider the influence of such molecular ions to be negligible for the following reasons.

First, considering the time from the chemical purification of $^{233}$U, the number of each descendant nuclide of $^{229}$Th and $^{232}$U extracted from the RF-carpet gas cell was at most ~10 ions/s. The ion-impact counts at −2000 V for such nuclides should be less than or comparable to ~10 counts/bunch, which is much smaller than those of $^{229\mathrm{m,g}}$Th$^{2+}$ (~$10^3$ counts/bunch, as shown in Fig. 3c). Hence, the effects of the descendant nuclides should be negligible for the $^{229\mathrm{m,g}}$Th$^{2+}$ data at −2000 V, even if molecular ions at *m*/*z* = 114.5 were present.

Second, for $^{229\mathrm{m,g}}$Th$^{+}$, the count rate at −2000 V (Fig. 3d) was an order of magnitude smaller than $^{229\mathrm{m,g}}$Th$^{2+}$. However, the number of molecular ions at *m*/*z* = 229.0 was estimated to be negligibly small. This is because the $\Gamma_{2+}$ values obtained from the $^{229\mathrm{m,g}}$Th$^{2+}$ and $^{229\mathrm{m,g}}$Th$^{+}$ data agreed with each other. If molecular ions had been significant in the $^{229\mathrm{m,g}}$Th$^{+}$ data, the obtained $\Gamma_{2+}$ value represents the production rate of the molecular ions, which should be generally different from the decay rate of $^{229\mathrm{m,g}}$Th$^{2+}$ ($\Gamma_{2+}$ value obtained from the $^{229\mathrm{m,g}}$Th$^{2+}$ data). Moreover, the most probable molecular ion at *m*/*z* = 229.0, namely, $^{213}$Bi$^{16}$O$^{+}$ would not have been produced in the ion trap because the reaction of $\mathrm{Bi^+ + NO \rightarrow BiO^+ + N}$ is endothermic at room temperature[40] due to the small bond energy of BiO$^{+}$ (1.8 eV), compared with that of NO (6.6 eV).

Third, the effects by molecular ions on the $^{229\mathrm{m,g}}$Th$^{2+}$ and $^{229\mathrm{m,g}}$Th$^{+}$ data at −35 V (Figs. 3e and 3f) should also be small, as discussed in sections "Isomer detection by IC electron measurement" and "Preparation of short ion bunches and IC-electron measurement". We also confirmed that the ratio of the MCP counts of $^{229\mathrm{m,g}}$Th$^{+}$ at −35 V to those at −2000 V was consistent between Campaigns A and B, where elapsed times from the $^{233}U$ purification were different; if molecular ions had significantly affected the results at −2000 or −35 V, such coincidence would not have been observed due to the growth of the descendant nuclides.

## Data Availability

The data supporting the findings of this study are available from the corresponding author upon request.

## Acknowledgements

We are sincerely grateful to F.F. Karpeshin for fruitful theoretical discussions. $^{233}U$ used in this study was provided by the $^{233}U$ cooperation project between the Japan Atomic Energy Agency and the Inter-University Cooperative Research Program of the Institute for Materials Research, Tohoku University (proposal numbers 17K0204, 17F0011, and 18F0014). This work was supported by JST (CREST Grant number JPMJCR24I6 to Y.S. and A.Y.) and JSPS (KAKENHI Grant Numbers JP17H01081 to H.H., JP19K23445 to Y.S., JP20K14500 to Y.S., JP21H04473 to Y.S., JP23K25904 to Y.S., JP23H00094 to A.Y. and Y.S., JP24H00469 to H.H. and Y.S., and JP25H00397 to Y.S.).

## Author contributions

Y.S., A.Y., K.T., and N.S developed the experimental apparatus. Y.S. and A.Y. performed the measurements and analyzed the data. Y.S., A.Y., H.K., K.S., and H.H. performed the chemical purification of $^{233}U$ and developed the $^{233}U$ source. M.W. developed the ion extracting system including the RF carpet and the quadrupole ion guide. All work was supervised by H.H, Y.K., and M.W. All authors discussed the results and contributed to the preparation of the manuscript.

**Competing interests**

The authors declare no competing interests.

**Correspondence and requests for materials** should be addressed to Yudai Shigekawa.

## Extended Data Figures

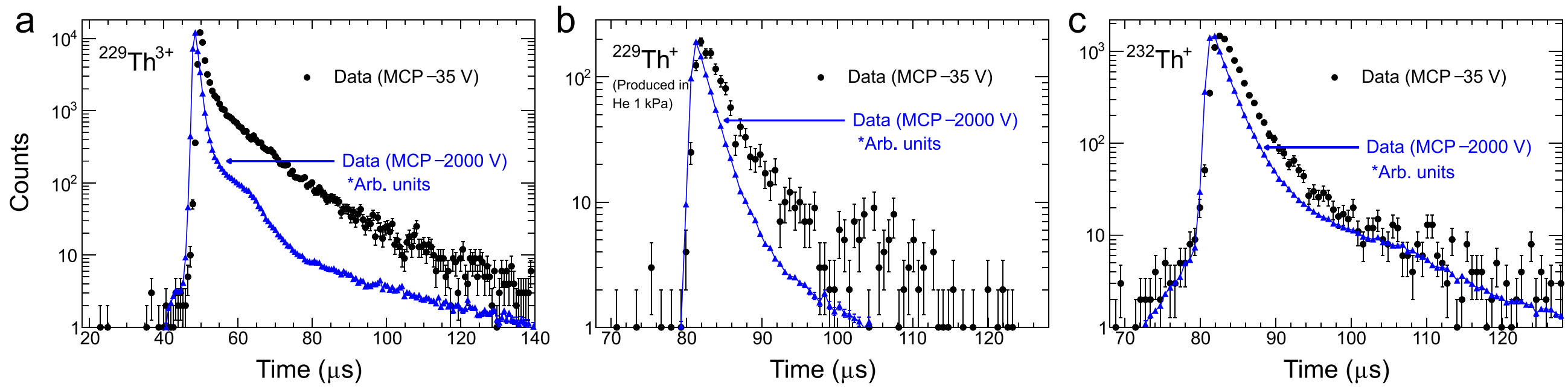

**Extended Data Fig. 1. Time traces of the MCP signals for the extraction of short ion bunches.** Black circles show MCP counts as a function of time (bin width 655 ns) at an MCP surface voltage of −35 V for $^{229m,g}Th^{3+}$ (a), $^{229m,g}Th^{+}$ (b), and $^{232}Th^{+}$ (c), while the blue triangles show the time traces obtained at an MCP surface voltage of −2000 V in arbitrary units. The blue solid curves are guides to the eye. The error bars in (a)–(c) represent 1 s.d. statistical uncertainties. The measurements for $^{229m,g}Th^{3+}$ and $^{229m,g}Th^{+}$ were performed in Campaign C, while those for $^{232}Th^{+}$ were performed in Campaign D. The $^{229m,g}Th^{+}$ ions in (b) were produced in the RF-carpet gas cell containing 1 kPa of He gas, not in the ion trap.

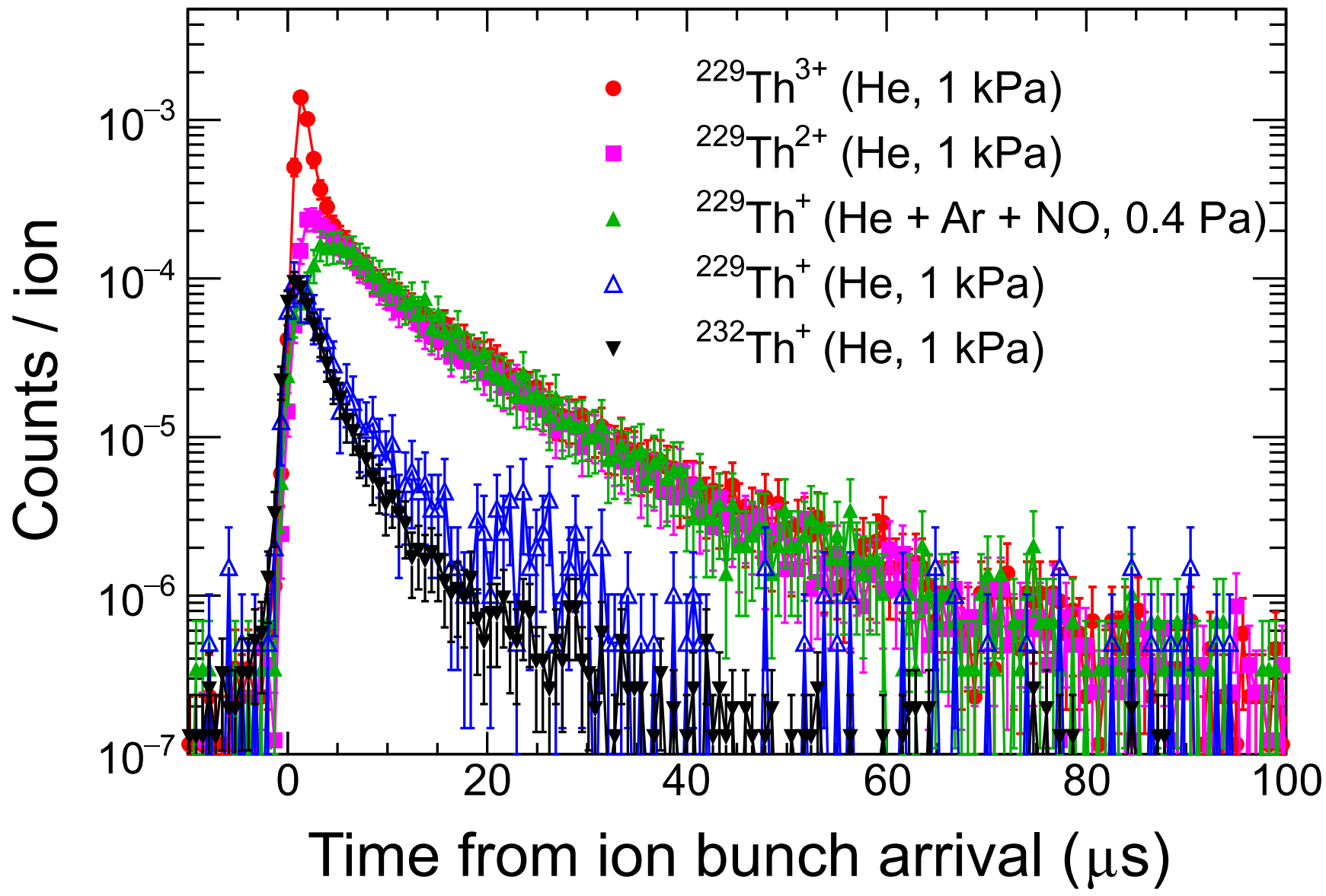

**Extended Data Fig. 2. Comparison of time traces of the MCP signals for the extraction of short ion bunches.** MCP counts at an MCP surface voltage of −35 V are divided by the total number of ions extracted on the MCP detector and shown as a function of time for the extraction of $^{229m,g}Th^{3+}$, $^{229m,g}Th^{2+}$, $^{229m,g}Th^{+}$, and $^{232}Th^{+}$ ions produced in the RF-carpet gas cell containing 1 kPa of He gas (red circle, purple square, blue open triangle, and black inverted triangle, respectively), as well as for $^{229m,g}Th^{+}$ ions produced by the charge-exchange reaction in the ion trap containing He, Ar, and NO gases at a total pressure of 0.40 Pa (green triangle). The 0 μs in this graph corresponds to the arrival time of the ion bunches at the MCP detector, measured at an MCP surface voltage of −2000 V. The solid curves are guides to the eye. The error bars represent 1 s.d. statistical uncertainties.

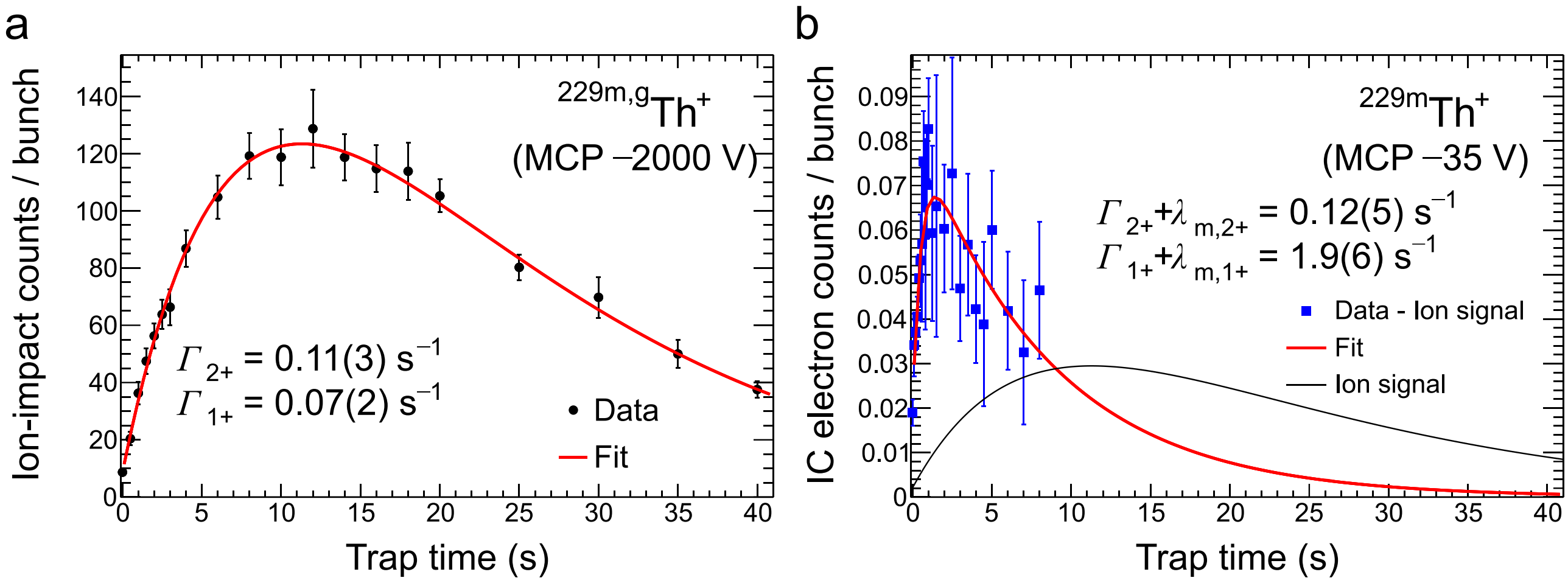

**Extended Data Fig. 3. Number of $^{229m,g}Th^{+}$ ions as a function of trapping time at a total pressure of 0.50 Pa. a**, Ion-impact counts (MCP counts at −2000 V) per ion bunch (black circle) as a function of the trapping time and the function (red curve) obtained by fitting Eq. (2) to the data. Here, the ion-impact counts of $^{229m,g}Th^{+}$ were regarded as those of $^{229g}Th^{+}$ due to the negligibly small fraction of $^{229m}Th$. **b**, IC-electron counts per ion bunch (blue square) as a function of the trapping time, and the function (red curve) obtained by fitting Eq. (4) to the data. IC-electron counts were obtained by subtracting the ion-impact counts (black curve) from the MCP counts at an MCP surface voltage of −35 V. The error bars in (a)–(b) include 1 s.d. statistical uncertainties and fluctuations in the number of $^{229}Th$ ions supplied by our RF carpet gas cell (5%). The gas pressures in the ion trap were kept at $p_{total}$ = 0.50 Pa, $p_{Ar}$ ≈ 0.38 Pa, and $p_{NO}$ = 4.6(6)×$10^{-7}$ Pa.

## Extended Data Tables

**Extended Data Table 1. Parameters derived from $^{229m,g}Th^{2+}$ measurements at different NO partial pressures.** Ion-impact counts per ion bunch as a function of the trapping time for $^{229m,g}Th^{2+}$ at each NO partial pressure ($p_{NO}$; first column) were fitted with Eq. (1), thereby yielding $\Gamma_{2+}$ (second column) for each $p_{NO}$. IC-electron counts per ion bunch as a function of the trapping time for $^{229m}Th^{2+}$ at each $p_{NO}$ were fitted with Eq. (3), thereby yielding $\Gamma_{2+} + \lambda_{m,2+}$ (third column) for each $p_{NO}$. The measurements were performed at a total pressure of 0.40 Pa. The $p_{NO}$ values in the table have systematic errors of 50% due to the uncertainty in the reaction rate constant[31] used to calculate $p_{NO}$.

| $p_{NO}$ (Pa) | $\Gamma_{2+}$ ($s^{-1}$) | $\Gamma_{2+} + \lambda_{m,2+}$ ($s^{-1}$) |
|---|---|---|
| $9.5(5)\times10^{-7}$ | 0.425(6) | 0.45(3) |
| $1.15(6)\times10^{-6}$ | 0.440(5) | 0.48(3) |
| $1.44(7)\times10^{-6}$ | 0.471(5) | 0.46(3) |
| $2.94(15)\times10^{-6}$ | 0.680(7) | 0.68(5) |

**Extended Data Table 2. Parameters derived from $^{229m,g}Th^+$ measurements at different NO partial pressures.** Ion-impact counts per ion bunch as a function of the trapping time for $^{229m,g}Th^+$ at each NO partial pressure ($p_{NO}$; first column) were fitted with Eq. (2), thereby yielding $\Gamma_{2+}$ (second column) and $\Gamma_{1+}$ (fourth column) for each $p_{NO}$. IC-electron counts per ion bunch as a function of the trapping time for $^{229m}Th^+$ at each $p_{NO}$ were fitted with Eq. (4), thereby yielding $\Gamma_{2+} + \lambda_{m,2+}$ (third column) and $\Gamma_{1+} + \lambda_{m,1+}$ (fifth column) for each $p_{NO}$. The $\lambda_{m,1+}$ value (sixth column) was obtained by subtracting $\Gamma_{1+}$ from $\Gamma_{1+} + \lambda_{m,1+}$ for each $p_{NO}$. The measurements were performed at a total pressure of 0.40 Pa. The $p_{NO}$ values in the table have systematic errors of 50% due to the uncertainty in the reaction rate constant[31] used to calculate $p_{NO}$.

| $p_{NO}$ (Pa) | $\Gamma_{2+}$ ($s^{-1}$) | $\Gamma_{2+} + \lambda_{m,2+}$ ($s^{-1}$) | $\Gamma_{1+}$ ($s^{-1}$) | $\Gamma_{1+} + \lambda_{m,1+}$ ($s^{-1}$) | $\lambda_{m,1+}$ ($s^{-1}$) |
|---|---|---|---|---|---|
| $9.5(5)\times10^{-7}$ | 0.44(6) | 0.40(5) | 0.26(2) | 1.8(4) | 1.5(4) |
| $1.15(6)\times10^{-6}$ | 0.43(8) | 0.42(5) | 0.27(3) | 2.0(4) | 1.7(4) |
| $1.44(7)\times10^{-6}$ | 0.42(8) | 0.45(5) | 0.30(4) | 1.9(4) | 1.6(4) |
| $2.94(15)\times10^{-6}$ | 0.57(10) | 0.69(7) | 0.48(7) | 1.7(3) | 1.3(4) |